\documentstyle[twocolumn,prl,aps]{revtex}
\begin{document}
\draft
\title{Systematic Inclusion of High-Order Multi-Spin Correlations\\
                for the Spin-$1\over2$ $XXZ$ Models}
\author{R. F. Bishop, R. G. Hale, and Y. Xian\cite{email}}
\address{Department of Mathematics,\\
         University of Manchester Institute
         of Science and Technology (UMIST),\\
         P.O. Box 88, Manchester M60 1QD, UK}
\date{\today}
\maketitle
\begin{abstract}
We apply the microscopic coupled-cluster method (CCM) to the
spin-$1\over2$ $XXZ$ models on both the one-dimensional chain and the
two-dimensional square lattice. Based on a systematic approximation
scheme of the CCM developed by us previously, we carry out high-order
{\it ab initio} calculations using computer-algebraic techniques. The
ground-state properties of the models are obtained with high accuracy
as functions of the anisotropy parameter. Furthermore, our CCM analysis
enables us to study their quantum critical behavior in a systematic
and unbiased manner.
\end{abstract}

%\begin{narrowtext}

Over the last few years the microscopic coupled-cluster method (CCM)
\cite{bikus} has successfully been applied to many lattice Hamiltonian
systems \cite{rohe,bpx,bkwx,rohes}, producing the best or among the
best results in terms of accuracy and power. For example, in the
so-called SUB2 approximation discussed below, the exact analytic
solution of the corresponding CCM equations often  exhibits a
terminating point. This is clearly demonstrated \cite{bpx} to
correspond to a physical critical point of the system by the behavior
within the same approximation of such calculated quantities as the
order parameter, correlation functions, and the gap in the excitation
spectrum. Recently we have developed several efficient, systematic
approximation schemes of the CCM specifically tailored for use with
lattice systems \cite{bpx,bkwx}.

In this article, we report our new results from an {\it ab initio}
high-order calculation for the spin-$1\over2$ $XXZ$ model on both the
one-dimensional chain (1DC) and the two-dimensional square lattice
(2DSL). In particular, we have obtained with high accuracy the
ground-state energy, the anisotropic susceptibility (i.e., the
second-order energy derivative with respect to the anisotropy
parameter), and the staggered magnetization, as functions of the
anisotropy parameter. Furthermore, since these physical quantities are
obtained as functions of the anisotropy parameter, we are able to
study the possible quantum phase transitions of the anisotropic models
in a systematic, unbiased manner. This is in contrast to the series
expansion technique \cite{seri} in which one has to make a Pad\'e
approximation or to assume a particular critical behaviour (from the
spin-wave theory of Anderson \cite{andes}, for example) for those
physical quantities from the outset.

Since the details of our CCM analysis have been published elsewhere
\cite{bpx}, we only outline the specific approximation schemes
employed here. The spin system under consideration is the so-called
spin-$1\over2$ $XXZ$ model described by the following Hamiltonian
\begin{equation}
  H={1\over2}\sum_{l=1}^N\sum_{\rho=1}^z
  \left[\Delta s_l^zs_{l+\rho}^z +
    {1\over2}(s_l^+s_{l+\rho}^- + s_l^-s_{l+\rho}^+)\right],
\end{equation}
where the index $l$ runs over all $N\ (\rightarrow \infty)$ lattice
sites with the usual periodic boundary condition imposed; the index
$\rho$ runs over all $z$ nearest-neighbour sites; the operators $s_l^z$
and $s_l^\pm\ (\equiv s_l^x \pm is_l^y)$ are spin operators; and
$\Delta$ is the anisotropy parameter. The special case $\Delta =1$
gives the isotropic Heisenberg model, which, with spin $s={1\over2}$ on
the 2DSL, has been under intensive study over the last six years or so
\cite{maba}.

We first consider the case of $\Delta \rightarrow \infty$. Equation
(1) then reduces to the Ising model with a classical ground state
(i.e., the N\'eel state) given by two alternating sublattices, one
with all spins down, the other with all spins up. For clarity, we use
the index $\{i\}$ exclusively for the spin-down sublattice, and the
index $\{j\}$ exclusively for the spin-up sublattice. Naturally, we
choose the N\'eel state as the model state $\vert\Phi\rangle$ in our
CCM analysis, and incorporate the quantum correlation effects by
considering the excitations with respect to this model state. The
elementary operators of these excitations are clearly given by the
spin-raising operators $s_i^+$ on the $i$-sublattice and the
spin-lowering operators $s_j^-$ on the $j$-sublattice. The CCM {\it
ansatz} for the ground ket-state is therefore given by
\begin{equation}
  \vert\Psi_g\rangle = {\rm e}^S\vert\Phi\rangle; \ \
  S \equiv \sum_{n=1}^{N/2} S_{2n},
\end{equation}
with the correlation operators $S_{2n}$ defined by
\begin{eqnarray}
  S_{2n} &=& {(-1)^n\over(n!)^2}\sum_{i_1,i_2,...,i_n}
  \sum_{j_1,j_2,...,j_n}{\cal S}_{i_1i_2...i_n,j_1j_2...j_n}
            \nonumber\\
  &&\times s_{i_1}^+s_{i_2}^+\cdots s_{i_n}^+
           s_{j_1}^-s_{j_2}^-\cdots s_{j_n}^-,
\end{eqnarray}
where we have restricted ourselves to the conserved sector of zero
$z$-component of total spin, $s_{\rm total}^z\ (\equiv \sum_{l=1}^N
s_l^z)$, by including only those configurations with equal numbers of
spin-flips on both sublattices.

The ground-state energy and the $c$-number coefficients $\{{\cal
S}_{i_1...i_n,j_1...j_n}\}$ of Eq.\ (3) are determined by taking the
inner products of the Schr\"odinger equation in the form ${\rm
e}^{-S}H{\rm e}^S\vert\Phi\rangle = E_g\vert\Phi\rangle$, firstly with
the model state itself and secondly with the states constructed by
acting on $\vert\Phi\rangle$ with the corresponding correlation
operators in $S_{2n}$. We thus find respectively the ground-state
energy $E_g$,
\begin{equation}
  E_g = \langle\Phi\vert{\rm e}^{-S}H{\rm e}^S\vert\Phi\rangle;
\end{equation}
and the coupled set of equations for $\{{\cal S}_{i_1...i_n,
j_1...j_n}\}$,
\begin{equation}
  \langle\Phi\vert s_{i_1}^-s_{i_2}^-\cdots s_{i_n}^-
  s_{j_1}^+s_{j_2}^+\cdots s_{j_n}^+ {\rm e}^{-S}H{\rm e}^S
      \vert\Phi\rangle = 0,
\end{equation}
with $n = 1,2,...,N/2$.

Each of the above equations always involves the Hamiltonian in a
similarity-transformed form, namely,
\begin{equation}
  {\rm e}^{-S}H{\rm e}^S =
  H + [H,S] + {1\over2!} [[H,S],S] + \cdots,
\end{equation}
where the expansion series terminates at the fourth order \cite{bpx}.
Therefore, once an approximation for $S$ is chosen, no further
approximation is necessary in order to obtain and solve the coupled set of
equations (5). For the spin-$1\over2$ $XXZ$ model of Eq.\ (1),
it is easy to derive the following exact equation for the ground state
energy per spin,
\begin{equation}
  {E_g\over N} = -{z\over8}(\Delta + 2 b_1),
\end{equation}
where $b_1 \equiv {\cal S}_{i,i+\rho}$ is the nearest-neighbour pair
correlation coefficient, and $z=2,4$ for the 1DC and 2DSL models
respectively. We note that $b_1$ is independent of both the
index $i$ and index $\rho$ by the lattice symmetries.

We clearly need an approximation method to truncate $S$ for any
practical calculation.  The three most commonly used truncation methods
are: the SUB$n$ scheme in which all correlations involving  only $n$ or
fewer spins are retained; the simpler SUB$n$-$m$ sub-approximation
scheme where only SUB$n$ correlations spanning a range of no more than
$m$ adjacent lattice sites are included; and finally the systematic
local LSUB$m$ scheme, which includes all possible multi-spin
correlations over a specified locale on the lattice, where $m$ is the
nominal index that characterizes the size of the given locale. In each
case the remaining correlation coefficients are set to zero. For
example, the LSUB4 scheme for the 1D spin-$1\over2$ model retains three
independent configurations, represented by $b_1,b_3$, and $g_4$
respectively \cite{bpx}. In particular, $b_1$ corresponds to the
nearest-neighbour two-spin-flip configuration mentioned before, $b_3$
to the third-nearest-neighbour two-spin-flip configuration, and $g_4$
to the spin-flip configuration of four adjacent spins. The
corresponding LSUB4 coupled equations are given by \cite{bpx},
\begin{eqnarray}
  &&1 -2\Delta b_1 -3b_1^2 +2b_1b_3+2b_3^2+2g_4 = 0,\\
  &&b_1^2-4\Delta b_3 -4b_1b_3+g_4 = 0, \\
  &&-\Delta(b_1^2+2b_1b_3)+g_4(\Delta+4b_1+b_3)+2b_1b_3^2 = 0.
\end{eqnarray}
After solving these coupled equations, we obtain the ground-state
energy by substituting $b_1$ into Eq.\ (7).

For the higher-order approximations the derivation of the coupled
equations becomes very tedious.  We have developed our own software
using C++ and Fortran to automate this process. Also, we have used
standard computer algebra packages to check our results independently.
For the LSUB$m$ schemes, we have derived and solved the coupled
equations up to $m=10$ for the 1DC case and up to $m=6$ for the 2DSL
case. It should be noted that all of our calculations were done on
microcomputers. The numbers of independent spin-flip configuration
coefficients retained in the these two cases are $81$ and $72$,
respectively.

Some of our results for the 1DC model have already been published
\cite{bpx}. In particular, we have shown that for a given value
of $m$ the LSUB$m$ scheme reproduces exactly the corresponding $2m$th
order of large-$\Delta$ perturbation theory, and that the LSUB$m$
scheme also gives good results in the planar region
($\vert\Delta\vert<1$) where perturbation theory is not valid. The new
high-order calculations further push the numerical results closer to
their exact counterparts obtained by the Bethe {\it ansatz}
\cite{beth}, over a wide range of $\Delta$ ($0 < \Delta < \infty$).
In particular, at the isotropic point ($\Delta=1$) the LSUB10 scheme
yields $-0.4420$ for the ground-state energy per particle. We have
made a naive attempt to extrapolate our results for the LSUB$m$ scheme
with $m=4,6,8,10$, and find that a $1/m^2$ rule seems to fit them
very well. The ground-state energy per particle after this
extrapolation yields $-0.4431\pm0.0001$ from a least squares fit,
while the exact result by Bethe {\it ansatz} \cite{beth} is $-0.4432$
to the accuracy of four significant figures.

In Fig.\ 1 we show some of our results for the ground-state energies
of the 2DSL model, together with a Monte Carlo calculation \cite{barn}
on an $8\times8$ lattice for comparison. One sees that the results for
our high-order calculations are in excellent agreement with the Monte
Carlo calculations over a wide range. At the isotropic point ($\Delta
=1$), our best numerical results for the ground-state energy of the
2DSL model, obtained from the LSUB6 scheme, is $-0.6670$. We find that
the $1/m^2$ rule also fits our 2D LSUB$m$ data well. The extrapolated
result is $-0.6691\pm0.0003$.  This number compares well with the
values $-0.66934\pm0.00004$ from a large-scale Monte Carlo calculation
by Runge \cite{MC}, and $-0.6694\pm0.0001$ from the series expansion
techniques \cite{seri}.

The more interesting feature from our 2DSL LSUB$m$ calculations is the
appearance of terminating points in the real solutions of the LSUB$m$
equations for $m>2$, as indicated in Fig. 1, beyond which the solution
becomes complex. This behavior is totally different from that observed
in the 1DC model where, unlike in the  SUB2 scheme discussed above
which retains two-spin correlations of arbitrarily long range, there is
no evidence of such terminating points in the localized LSUB$m$ scheme
at any value $m \le 10$. Since we have derived the corresponding
coupled equations for a given LSUB$m$ scheme as closed analytical forms
with $\Delta$ as a parameter, we can straightforwardly study the
terminating points by taking the derivatives on both sides of the
coupled set of equations (e.g., Eqs.\ (8-10)) with respect to $\Delta$
and solving directly for the derivatives of the coefficients. From Eq.\
(7), we define the second-order derivative of the ground-state energy
as the anisotropic susceptibility $\chi_a$,
\begin{equation}
  \chi_a \equiv -{\partial^2 (E_g/N)\over \partial\Delta^2} =
  {z\over4}{\partial^2 b_1\over\partial\Delta^2}.
\end{equation}
The numerical results for $\chi_a$ as a function of $\Delta$ are shown
in Fig.\ 2 for the 1DC and 2DSL models respectively. It is clear from
the figure that there is no singular behavior for the 1DC model. By
comparison the exact calculation \cite{beth} gives an essential
singularity at $\Delta=1$, with the result that any finite order of
derivative with respect to $\Delta$ is indeed continuous. However, the
anisotropic susceptibility for the 2DSL model clearly shows a singular
behavior (except for the low-order LSUB2 scheme).  Furthermore,
although the values of the terminating points $\Delta_m$ are different
for the LSUB4 and LSUB6 cases, both schemes yield similar critical
behavior, namely
\begin{equation}
  \chi_a \rightarrow {\rm const.}\times
  (1 - x^2)^{-\lambda}, \ \ \ \ x \rightarrow 1,
\end{equation}
with $\lambda = 3/2$, and $x \equiv \Delta_m/\Delta$. This clearly
suggests that $E_g/N$ for the 2DSL model within the LSUB$m$
approximation has the following expansion near the terminating point,
\begin{eqnarray}
  {E_g\over N} &\rightarrow& A_m + B_m(1-x^2)^{1/2}
      +C_m(1-x^2)\nonumber\\
  &&+ D_m(1-x^2)^{3/2} + \cdots,
\end{eqnarray}
where $x \equiv \Delta_m/ \Delta$.

We note that the spin-wave theory of Anderson \cite{andes} gives the
exponent $\lambda =1/2$. It is not difficult to show that our full
SUB2 scheme also yields the same value \cite{bpx}, whereas from the
above we see that the LSUB$m$ schemes yield $\lambda = 3/2$. In order
to understand this difference, we consider the SUB2-$m$ scheme of the
2DSL model with $m \le 14$. We observe that the SUB2-$m$ scheme also
has the singular behavior described by Eq.\ (13), yielding $\lambda =
3/2$. However, the coefficient $B_m \rightarrow 0$ as
$m\rightarrow\infty$. Therefore the fourth term in Eq.\ (13) is the
dominant singular term, yielding $\lambda =1/2$ for the full SUB2
scheme (i.e., the SUB2-$\infty$ scheme).  Furthermore, we notice that
the coefficients $B_m$ also seems to decay by a $1/m^2$ rule as $m$
increases in the SUB2-$m$ scheme. It seems likely that in the LSUB$m$
scheme, the $B_m$ coefficients also decrease as $m$ increases, and the
first singular term has the coefficient $D_m$ as $m\rightarrow
\infty$, so that $\lambda=1/2$ is recovered. Although we only have
the LSUB4 and LSUB6 schemes for consideration, the results confirm
that the value of $B_m$ in the LSUB6 scheme is much smaller than that
of the LSUB4 scheme.  Therefore we expect that the exact result,
namely the LSUB$m$ scheme with $m\rightarrow \infty$, has the value of
$\lambda = 1/2$. We will present the details of the expansion Eq.\
(13) for our various approximation schemes elsewhere.

While our CCM analysis seems to agree with spin-wave theory
\cite{andes} for the critical exponent $\lambda$, there does seem to be
a real difference in the corresponding predictions for the value of the
critical point $\Delta_c$. We firstly consider the SUB2-$m$ scheme. As
given previously \cite{bpx}, the terminating point is
$\Delta_\infty=0.7984$. The terminating point in the SUB2-$m$ scheme
rapidly approaches this value as $m$ increases. Interestingly, the
$1/m^2$ rule also seems to fit well. We attempt to apply the $1/m^2$
rule for the LSUB$m$ scheme also. We thus obtain the predicted critical
point as $\Delta_c \approx 0.92\pm0.01$. Although this is clearly
smaller than the value 1 predicted by spin-wave theory \cite{andes},
the quoted error is merely that of least-squares fit at this level, and
we cannot yet preclude agreement when higher-order corrections in the
LSUB$m$ scheme (i.e., $m>6$) are included in the extrapolation.

Finally, we calculate the staggered (sublattice) magnetization $M^z
\equiv \langle s_j^z \rangle/s$ for the $XXZ$ models, where the
expectation value is taken with respect to the ground ket and bra
states within the same CCM approximation schemes \cite{bpx}.  Our
results show, as expected, that there is clearly a difference between
the 1DC and 2DSL models. For the 1DC case, the staggered magnetization
shows no singular behavior at all, even for the LSUB10 scheme. However,
for the 2DSL model, the singular behavior is clearly seen for all
high-order schemes beyond the LSUB2 scheme. For the 2DSL model, at the
isotropic point, $M^z = 0.8514, 0.7648, 0.7278$ in the LSUB2, LSUB4 and
LSUB6 schemes respectively.  From these numbers we obtain $M^z =
0.68\pm0.01$ by extrapolation using a $1/m$ rule (which was found to
work well for the SUB2-$m$ scheme). This value is somewhat bigger than
the corresponding values $0.606$ from spin-wave theory \cite{andes},
and $0.62\pm0.02$ from series expansion techniques \cite{seri},
although it agrees well with the best of the corresponding Monte Carlo
results, which vary between $0.68\pm0.02$ and $0.62\pm0.04$ \cite{MC}.

In conclusion, our high-order CCM analyses not only produce with high
accuracy the ground-state properties for the spin-$1\over2$ $XXZ$
models, but they also enable us to study the the quantum phase
transitions in a systematic, unbiased manner. They clearly show that
the LSUB$m$ approximations themselves represent a natural extension of
perturbation theory. In effect, they comprise a well-defined analytical
continuation or resummation of the perturbation theory results within
the context of a natural and consistent hierarchy of approximations.
More significantly, this hierarchy is capable of predicting a possible
quantum phase transition. Another calculation under consideration is
the energy gap of the excited states near the critical point in the
LSUB$m$ scheme. Our result \cite{bpx} for the energy gap in the SUB2
scheme is quite similar to that of spin-wave theory, but higher-order
multi-spin correlations have been proved to be significant
\cite{seri,barn2}.

Finally, we note that the combination of the CCM as a theoretical
framework and the use of computer-algebraic techniques to implement it
at high orders of approximation, has resulted in a formalism which is
capable for the 2D spin-lattice models of attaining numerical
results with a precision comparable to those from the much more
computationally intensive state-of-the-art Monte Carlo simulations. It
will be of great interest to apply our techniques to similar
electron-lattice problems of interest in high-temperature
superconductivity which involve vacancies on the lattice (e.g., the
Hubbard model), where Monte Carlo algorithms are not easily
applicable, due to the infamous fermion sign problem.

\acknowledgments
We are grateful to J. B. Parkinson and C. Zeng for many useful
discussions. One of us (R. F. B.) also gratefully acknowledges a
research grant from the Science and Engineering Research Council
(SERC) of Great Britain.

\begin{figure}
\caption{Ground-state energy per spin as a function of $\Delta$ for the
spin-$1\over2$ $XXZ$ model on the 2DSL. Shown are the numerical
results of the LSUB$m$ scheme with $m=2,4,6$ and of a Monte
Carlo calculation of Ref. [10].}
\end{figure}
\begin{figure}
\caption{The second-order derivative of the ground-state energy
per spin with respect to $\Delta$ for the 1DC and 2DSL models as
functions of $\Delta$. Shown are the results of several LSUB$m$
schemes.}
\end{figure}

%\end{narrowtext}

\begin{references}
\bibitem[\dag]{email}xian@lanczos.ma.umist.ac.uk
\bibitem{bikus}R. F. Bishop and H. K\"ummel, Phys. Today {\bf 40(3)},
52 (1987); R. F. Bishop, Theor. Chim. Acta {\bf 80}, 95 (1991).
\bibitem{rohe}M. Roger and J. H. Hetherington, Phys. Rev. B {\bf 41},
200 (1990); F. E. Harris, Phys. Rev. B {\bf 47}, 7903 (1993).
\bibitem{bpx}R. F. Bishop, J. B. Parkinson, and Y. Xian, Phys. Rev. B
{\bf 43}, 13782 (1991); {\bf 44}, 9425 (1991); J. Phys.: Condens.
Matter {\bf 4}, 5783 (1992).
\bibitem{bkwx}R. F. Bishop, A. S. Kendall, L. Y. Wong, and Y. Xian,
Phys. Rev. D {\bf 48}, 887 (1993); R. F. Bishop and Y. Xian, Acta
Phys. Pol. B {\bf 24}, 541 (1993).
\bibitem{rohes}M. Roger and J. H. Hetherington, Europhys. Lett. {\bf
11}, 255 (1990); C. F. Lo, E. Manousakis, and Y. L. Wang, Phys. Lett. A
{\bf 156}, 42 (1991); F. Petit and M. Roger, Phys. Rev. B {\bf 49}, 3453
(1994).
\bibitem{seri}R. R. P. Singh, Phys. Rev. B {\bf 39}, 9760 (1989); R.
R. P. Singh and D. A. Huse, {\it ibid.} {\bf 40}, 7247 (1989); W.
Zheng, J.  Oitmaa, and C. J. Hamer, {\it ibid.} {\bf 43}, 8321 (1991).
\bibitem{andes}P. W. Anderson, Phys. Rev. {\bf 86}, 694 (1952);
T. Oguchi, {\it ibid.} {\bf 117}, 117 (1960).
\bibitem{maba}E. Manousakis, Rev. Mod. Phys. {\bf 63}, 1 (1991); T.
Barnes, Int. J. Mod. Phys. B {\bf2}, 659 (1991).
\bibitem{beth}H. A. Bethe, Z. Phys. {\bf 71}, 205 (1931); L.
Hulth\'en, Ark. Mat. Astron. Fys. A {\bf 26}, No. 11 (1938).
\bibitem{barn}T. Barnes, D. Kotchan, and E. S. Swanson, Phys. Rev. B
{\bf 39}, 4357 (1989).
\bibitem{MC}J. Carlson, Phys. Rev. B {\bf40}, 846 (1989); N. Trivedi
and D. M. Ceperley, {\it ibid.} {\bf 41}, 4552 (1990); K. J. Runge, {\it
ibid.} {\bf 45}, 12292 (1992).
\bibitem{barn2}T. Barnes {\it et al.}, Phys. Rev. B {\bf 40}, 8945
(1989).

\end{references}
\end{document}